\documentclass[a4paper,11pt]{article}

\usepackage[margin=1in]{geometry}
\usepackage{fullpage}
\usepackage{amsthm}
\usepackage{amsfonts}
\usepackage{amssymb}
\usepackage{amsmath,amsthm}
\usepackage{multirow}
\usepackage{color}
\usepackage{xspace}
\usepackage{clrscode}

\usepackage{ifpdf}
\ifpdf    
\usepackage{hyperref}
\else    
\usepackage[hypertex]{hyperref}
\fi
\allowdisplaybreaks

\newtheorem{theorem}{Theorem}[section]

\newtheorem{conjecture}{Conjecture}[section]

\newcommand{\comment}[1]{}
\newcommand{\junk}[1]{}

\newcommand{\ThreeSUMIndexing}{\textsf{3SUM-Indexing}}
\newcommand{\KSUMIndexing}{\textsf{KSUM-Indexing}}
\newcommand{\ThreeSUM}{\textsf{3SUM}}
\newcommand{\KSUM}{\textsf{KSUM}}

\newcommand{\ignore}[1]{}
\newcommand{\remove}[1]{}
\newcommand{\reals}{\hbox{$\rlap{\rm I} \> \kern-.2mm{\rm R}$}}

\usepackage{authblk}

\begin{document}
{
\title{The Strong 3SUM-INDEXING Conjecture is False}

\author[ ]{Tsvi Kopelowitz}
\author[ ]{Ely Porat}
\affil[ ]{Bar-Ilan University, Ramat Gan, Israel}
\affil[ ]{\texttt{kopelot@gmail.com, porately@cs.biu.ac.il}}

\date{}

\maketitle
\thispagestyle{empty}
\setcounter{page}{0}

\begin{abstract}
In the \ThreeSUMIndexing{} problem the goal is to preprocess two lists of elements from $U$,  $A=(a_1,a_2,\ldots,a_n)$ and $B=(b_1,b_2,...,b_n)$, such that given an element $c\in U$ one can quickly determine whether there exists a pair $(a,b)\in A \times B$ where $a+b=c$.
Goldstein et al.~[WADS'2017] conjectured that there is no algorithm for \ThreeSUMIndexing{} which uses $n^{2-\Omega(1)}$ space and $n^{1-\Omega(1)}$ query time. 

We show that the conjecture is false by reducing the \ThreeSUMIndexing{} problem to the problem of inverting functions, and then applying an algorithm of Fiat and Naor [SICOMP'1999] for inverting functions.
\end{abstract}

\newpage
\section{Introduction}\label{sec:intro}
In the \ThreeSUM{} problem the input is three sets $A$, $B$ and $C$, each containing $n$ elements from a universe $U$ that is closed under addition, and the goal is to establish whether there exists a triplet $(a,b,c)\in A\times B\times C$ such that $a+b=c$.
The \ThreeSUM{} conjecture, which states that there is no algorithm in the RAM model that solves \ThreeSUM{} in $n^{2-\Omega(1)}$ time~\cite{GajentaanO95,KPP16}, is one of the most popular conjectures used for proving conditional lower bounds on the time cost of various algorithmic problems~\cite{Patrascu10,WW13,AWW14,ACLL14,AW14,AWY15,KPP15,AKLPPS16,GKLP16,GKLP17}.

Together with Goldstein and Lewenstein in~\cite{GKLP17}, we considered an \emph{online} variant of the \ThreeSUM{} problem, which we called the
\ThreeSUMIndexing{} problem.
In the \ThreeSUMIndexing{} problem the goal is to preprocess two lists of elements from $U$,  $A=(a_1,a_2,\ldots,a_n)$ and $B=(b_1,b_2,...,b_n)$, such that given an element $c\in U$ one can quickly determine whether there exists a pair $(a,b)\in A \times B$ where $a+b=c$.
One straightforward algorithm for solving the \ThreeSUMIndexing{} problem is to store the sums of all pairs of values in $A \times B$. The space usage of this algorithm is $O(n^2)$ and the query time is $O(1)$ (using a hash table).
Another straightforward algorithm is to separately sort $A$ and $B$, and answer a query in $O(n)$ time by scanning $A$ forward and $B$ backwards. The space usage of this second algorithm is $O(n)$ words.
In~\cite{GKLP17} we argued that it is unclear how one can do better than these algorithms, which led us to introduce the following conjecture regarding the \ThreeSUMIndexing{} problem.
\begin{conjecture}
\textbf{Strong \ThreeSUMIndexing{} Conjecture}: There is no solution for the \ThreeSUMIndexing{} problem with $n^{2-\Omega(1)}$ space and $n^{1-\Omega(1)}$  query time.
\end{conjecture}

\paragraph{Our results.}
In this paper we design an algorithm for the \ThreeSUMIndexing{} problem, which is summarized in the following theorem, and refutes the strong \ThreeSUMIndexing{} conjecture.

\begin{theorem}\label{thm:alg3SUMINDEX}
For any $0<\delta<1$ there exists an algorithm that solves the \ThreeSUMIndexing{} problem whose space usage is $O(n^{2-\delta/3})$ words and the cost of a query is $O(n^{\delta})$ time.
\end{theorem}

Notice that if $\delta = 3/4$ then the space usage is $n^{1.75}$ while the query time is $n^{0.75}$, thereby refuting the \ThreeSUMIndexing{} conjecture.

\paragraph{Related work.}
Independently from our work, Golovnev et al.~\cite{GGHPV19} discovered a similar algorithm for the \ThreeSUMIndexing{} problem.

\section{The Algorithm}
\paragraph{\KSUMIndexing{}.}
To prove Theorem~\ref{thm:alg3SUMINDEX} we prove a more general theorem for the \KSUMIndexing{} problem, where the goal is to preprocess $k-1$ lists of elements $A_1,A_2,\ldots,A_{k-1}\in U^n$ where $A_i = (a_{i,1}, a_{i,2}, \ldots, a_{i,n})$, such that given an element $c\in U$ one can quickly determine whether there exists $(x_1,x_2,\ldots,x_{k-1})\in A_1\times A_2 \times \cdots \times A_{k-1}$ such that $\sum_{i=1}^{k-1}x_i=c$.
Notice that Theorem~\ref{thm:algKSUMINDEX} implies Theorem~\ref{thm:alg3SUMINDEX} when $k=3$.

\begin{theorem}\label{thm:algKSUMINDEX}
For any $0<\delta<1$ and a constant natural $k$, there exists an algorithm that solves the \KSUMIndexing{} problem whose space usage is $O(n^{k-1-\delta/3})$ words and the cost of a query is $O(n^{\delta})$ time.
\end{theorem}

The algorithm has two main ingredients:
a carefully constructed function $f:[n]^{k-1}\rightarrow [n]^{k-1}$
and a data structure for inverting $f$.

Let $A_1+A_2+\cdots+A_{k-1}= \big\{\sum_{j=1}^{k-1}x_j|(x_1,x_2,\ldots,x_{k-1})\in A_1\times A_2\times\cdots\times A_{k-1}\big\}.$

\paragraph{The function $f$.}
Let $g:[n]^{k-1}\rightarrow U$ be defined as $g(i_1,i_2,\ldots,i_{k-1}) = \sum_{j=1}^{k-1}a_{j,i_j}$.
Let $h:U\rightarrow [n]^{k-1}$ be a function which maps elements from $U$ to $k-1$ indices.
The function $f$ is defined as $f(i_1,i_2,\ldots,i_{k-1}) = h\big(g(i_1,i_2,\ldots,i_{k-1})\big)$.
The role of $f$ is to provide a mechanism for efficiently moving from $k$ indices $(i_1,i_2,\ldots,i_{k-1})$ to $k$ indices $(i'_1,i'_2,\ldots,i'_{k-1}) = f(i_1,i_2,\ldots,i_{k-1})$.

\paragraph{Inverting $f$.}
The key observation that the algorithm leverages is that when given $c\in U$ during query time, if there exist $k-1$ indices $(i_1,i_2,\ldots,i_{k-1}) \in A_1\times A_2\times \cdots \times A_{k-1}$ such that $\sum_{j=1}^{k-1}a_{j,i_j} = c$, then $$f(i_1,i_2,\ldots,i_{k-1}) = h(g(i_1,i_2,\ldots,i_{k-1})) = h(\sum_{j=1}^{k-1}a_{j,i_j}) = h(c).$$
Thus, if the answer to the \KSUMIndexing{} query $c$ is ``yes" then there exists $i_1,i_2,\ldots,i_{k-1}\in [n]$ such that $ (i_1,i_2,\ldots,i_{k-1}) \in f^{-1} (h(c))$.

In order to compute $f^{-1}$, the algorithm uses a result by Fiat and Naor~\cite{FiatN99} which is summarized by the following theorem\footnote{We remark that the algorithm of Fiat and Naor in~\cite{FiatN99} is randomized, while the version of the algorithm stated in Theorem~\ref{thm:FN} is deterministic. The difference is due to the unlimited preprocessing time which allows to convert the algorithm of~\cite{FiatN99} into a deterministic algorithm.}.

\begin{theorem}\label{thm:FN}
  For any function $f:D\rightarrow D$ where $|D|= N$ and for any choice of values  $(S,T)$ such that $T\cdot S^3 = N^3$, there exists an algorithm for inverting $f$ that uses $\tilde O(S)$ words of space and inverts $f$ in $\tilde O(T)$ time.
\end{theorem}

In our setting, $N=n^{k-1}$, and so if the query time is $T=\tilde \Theta(n^{\delta})$ for some $0<\delta <1$, then the space usage can be chosen to be  $$S=\left(\frac{N^3}{T}\right)^{\frac 13} = \tilde \Theta\left(n^{k-1-\frac{\delta} 3}\right) .$$

\paragraph{Dealing with large pre-images.}
The inversion algorithm of Fiat and Naor returns only one of the elements in $f^{-1}$.
Thus, if the Fiat and Naor algorithm returns $(i_1,i_2,\ldots,i_{k-1}) \in f^{-1} (h(c))$, we cannot guarantee that $\sum_{j=1}^{k-1}a_{j,i_j} = c$ even if the answer to the query is ``yes".
Nevertheless, notice that if $h$ is chosen so that for any $z,z'\in A_1+A_2+\cdots+A_{k-1}$, we have $h(z)\ne h(z')$, then we are guaranteed that if the answer to the query is ``yes" then the size of $f^{-1}(c)$ is one\footnote{Notice that there could be several elements in $[n]^k$ that map through $g$ to some $c\in U$. However, for the purpose of solving \KSUMIndexing{} it suffices to consider only one of those elements.}.
  Unfortunately, in general storing such a function $h$ requires too much space.
So instead, suppose $h$ is a pair-wise independent function, and so for $x,y\in U$ where $x\ne y$ we have $\Pr[h(x)=h(y)]=\frac{1}{n^{k-1}}$.
Moreover, the function $h$ is stored in $O(k)=O(1)$ words of space.
An element $z\in A_1+A_2+\cdots+A_{k-1}$ is said to be \emph{$h$-singleton} if for any $z'\in A_1+A_2+\cdots+A_{k-1}$ where $z\ne z$ we have $h(z)\ne h(z')$.
From the properties of pair-wise independent functions, there are $\Omega(n^{k-1})$ $h$-singleton elements.
Thus, if $c$ is $h$-singleton then there is only one element in $f^{-1}(c)$.

Now, if instead of choosing one $h$ function the algorithm repeats the process $O(\log n)$ times with functions $h_1,h_2\ldots,h_{O(\log n)}$, then with high probability each $z\in A_1+A_2+\cdots+A_{k-1}$ is a $h_\ell$-singleton for some $1\le \ell\le O(\log n)$. Notice that since the algorithm is allowed to have a large preprocessing time, the algorithm can select the functions  $h_1,h_2\ldots,h_{O(\log n)}$ while deterministically guaranteeing that each $z\in A_1+A_2+\cdots+A_{k-1}$ is a $h_\ell$-singleton for some $\ell\le O(\log n)$.

Finally, the algorithm constructs $f_\ell = h_\ell(g(i_1,i_2,\ldots,i_{k-1}))$ for $1\le \ell \le O(\log n)$, and for each $f_\ell$ the algorithm preprocesses $f_\ell$ using the Fiat and Naor algorithm.
Given a query $c$, the algorithm computes $f_\ell^{-1}(c)$ for all $1\le \ell \le O(\log n)$, and for each
$(i_1,i_2,\ldots,i_{k-1})=f_\ell^{-1}(c)$ the algorithm computes $g(i_1,i_2,\ldots,i_{k-1})$ in $O(k)=O(1)$ time. If $g(i_1,i_2,\ldots,i_{k-1})=c$ at least once, then the algorithm returns ``yes".
However, if $g(i_1,i_2,\ldots,i_{k-1})\ne c$ for all $1\le \ell \le O(\log n)$ then the algorithm answers ``no".

Finally, using $O(\log n)$ functions implies that if the query time is $T=\tilde O(n^\delta)$ then the space usage is $S=\tilde O(n^{k-1-\delta/3})$.

\subsection{Random Instances of \KSUMIndexing{}}
For the \KSUM{} problem, even random inputs are considered to be hard.
However, for \KSUMIndexing{}, our algorithm obtains faster bounds for random instances.
The reason for this improvement is that the tradeoff cost of the algorithm of Fiat and Naor~\cite{FiatN99} is reduced to $T\cdot S^2 = N^2$ whenever the function $f$ is random\footnote{See the discussion in Section 3 of~\cite{FiatN99}.}, and since in our case $f(i_1,i_2,\ldots,i_{k-1})$ depends on the values of $a_{1,i_1},a_{2,i_2},\ldots, a_{k-1,i_{k-1}}$ which are all random elements, we obtain the following theorem.

\begin{theorem}
For any $0<\delta<1$ there exists an algorithm that solves the \ThreeSUMIndexing{} problem on a uniformly random instance whose space usage is $O(n^{k-1-\delta/2})$ words and the cost of a query is $O(n^{\delta})$ time.
\end{theorem}

Notice that, for \ThreeSUMIndexing{}, if $\delta = 3/4$ then the space usage of the algorithm for random instances  is $n^{1.625}$ while the query time is $n^{0.75}$.

\section{Acknowledgments}
This research is supported by ISF grants no. 824/17 and 1278/16 and by an ERC grant MPM under the EU's Horizon 2020 Research and Innovation Programme (grant no. 683064).

\bibliographystyle{plain}
\bibliography{tsvi,ref}

\end{document}